\documentclass[12pt]{article}
\input epsf.tex
\usepackage{graphicx}
\makeatletter
\@addtoreset{equation}{section}

\newcommand{\be}{\begin{equation}}
\newcommand{\ee}{\end{equation}}
\newcommand{\bear}{\begin{eqnarray}}
\newcommand{\eear}{\end{eqnarray}}
\newcommand{\ba}{\begin{array}}
\newcommand{\ea}{\end{array}}

\newcommand{\CL}{{\cal L}} \newcommand{\CE}{{\cal E}}
\newcommand{\CN}{{\cal N}} 
\newcommand{\CH}{{\cal H}} 
 \newcommand{\mn}{{\mu\nu}}
\newcommand{\p}{{\partial}} \newcommand{\ra}{{\rightarrow}}
\newcommand{\CP}{{\cal P}} \newcommand{\del}{{\delta}}
\newcommand{\vX}{{\vec{X}}} \newcommand{\dvX}{{\del \vec{X}}}
\newcommand{\dXtau}{{\delta\vec{X}_\tau}}
\newcommand{\dXprime}{{\delta\vec{X}'}}
\newcommand{\nhat}{{\hat{n}}}
\newcommand{\thetahat}{{\hat{\theta}}}

\begin{document}

\begin{titlepage}
\vfill
\begin{flushright}
{\normalsize KIAS-P04008}\\
{\normalsize hep-th/0402027}\\
\end{flushright}

\vfill
\begin{center}
{\Large\bf Open/Closed Duality, Unstable D-Branes \\ and Coarse-Grained
Closed Strings }

\vskip 0.3in

{ Ho-Ung Yee\footnote{\tt ho-ung.yee@kias.re.kr} and Piljin
Yi\footnote{{\tt piljin@kias.re.kr}} }

\vskip 0.15in

 {\it School of Physics, Korea Institute for Advanced Study,} \\
{\it 207-43, Cheongryangri-Dong, Dongdaemun-Gu, Seoul 130-722, Korea}
\\[0.3in]

{\normalsize  2003}

\end{center}

\vfill

\begin{abstract}
\normalsize\noindent At the final stage of unstable D-brane decay
in the effective field theory approach, all energy and momentum of
the initial state are taken up by two types of fluids, known as
string fluid and tachyon matter. In this note, we compare motion
of this fluid system to that of macroscopic collection of
stretched closed strings and find a precise match at classical
level. The string fluid reflects low frequency undulation of the
stretched strings while the tachyon matter encodes the average
effect of high frequency oscillations turned on those strings. In
particular, the combined fluid system has been known to have a
reduced speed of light, depending on the composition, and we show
that this property is exactly reproduced in classical motion on
the closed string side. Finally we illustrate how the tachyon
matter may be viewed as an effective degrees of freedom carrying
high frequency energy-momentum of Nambu-Goto strings by
coarse-graining the dynamics of the latter.

\end{abstract}

\vfill

\end{titlepage}
\setcounter{footnote}{0}

\baselineskip 18pt
\pagebreak
\renewcommand{\thepage}{\arabic{page}}
%\tableofcontents
\pagebreak

\section{Introduction}

The low energy approach \cite{Sen:1998sm} to unstable D-branes has
proved to be surprisingly rich and successful. Regardless of
whether we start with Born-Infeld type action
\cite{Sen:1999xm,Sen:1999md,Garousi:2000tr,Bergshoeff:2000dq,
Kluson:2000iy,Kutasov:2003er} or with that from boundary string
field theory (BSFT)
\cite{Gerasimov:2000zp,Kutasov:2000qp,Kutasov:2000aq,Kraus:2000nj,Takayanagi:2000rz},
it successfully reproduced lower dimensional D-brane as
worldvolume solitons with exactly the right tension. Furthermore,
starting with Born-Infeld type action of unstable D-brane, the
correct low energy effective action of those stable D-branes was
obtained \cite{Sen:2003tm,Arutyunov:2000pe}. All these happen
despite the fact that the open string theory is severely truncated
for the low energy approximation. For some reason, the latter
remembers quite a lot of what the full string theory is
expected to have.

Another interesting aspect of the low energy theory is found in the
sector with net electric flux
\cite{Bergman:2000xf,Gibbons:2000hf}. The gauge field on
the unstable D-brane can have conserved electric flux, and these
flux lines carry fundamental string charges \cite{Yi:1999hd}. When the D-brane
decay reaches its final stage, the classical solution of the
system is characterized as a two component fluid system
\cite{Gibbons:2002tv,Kwon:2003qn}. One is the pressureless
electric flux lines, dubbed as string fluid \cite{Gibbons:2000hf},
while the other is dust-like object known as tachyon matter
\cite{Sen:2002nu,Sen:2002in,Sen:2002an}. Both fluid components
behave as perfect fluid, albeit of different dimensionality.

Most general static solution of unstable D-brane decay has been
found, and can be described in a simple term
\cite{Gibbons:2000hf,Gibbons:2002tv,Kwon:2003qn}: Flux lines
should be all collinear but otherwise unrestricted in their
distribution. Tachyon matter can be distributed arbitrarily as
well, except that its density should be uniform along the flux
line direction. This simple state comes with huge degeneracy
because transverse (with respect to flux lines) distribution of
neither fluid density affects the total energy.\footnote{A simpler
version of such configurations, where both fluids are taken to be
uniform, is also realized as a boundary state in string theory
\cite{Sen:2002nu,Mukhopadhyay:2002en} and further studied in
Ref.~\cite{Rey:2003xs}, although it is very difficult to see
dynamics of the final stage from this approach.}

In the prime motivation for studying this case with net electric
flux is a fundamental question on the relationship between open
strings and closed strings. The tachyon condensation is supposed
to remove all open string degrees of freedom, and the final state
must be described in terms of closed strings only. {}From this
viewpoint, the string fluid and tachyon matter must also have a
natural interpretation via closed string states. On the other
hand, the low energy description of the system is tree-level in
the open string sense, where closed string is usually not visible.
Thus, if we indeed succeed in describing these fluid in terms of
closed strings, it begs for further explanations of how the two
kinds of strings are related.

The string fluid has been known for some time by now to mimic
classical behavior of fundamental string remarkably well
\cite{Gibbons:2000hf}. In particular, one could start with a
configuration where a finite amount of flux lines are bundled into
an infinitesimal tubular shape. Dynamics of such a configuration
has been shown to be exactly that of a Nambu-Goto string
\cite{Gibbons:2000hf,Sen:2000kd,Nielsen} whose tension is nothing
but the total flux. Natural construction of fundamental strings
from this is, however, hampered by the degeneracy of the string
fluid. No confinement mechanism is known within the classical
regime, while a fully quantum scenario has been suggested
\cite{Bergman:2000xf,Yi:1999hd}.

More recently, a macroscopic interpretation for the combined
system of string fluid and tachyon matter is proposed
\cite{Sen:2003xs}. Consider a macroscopic number of long
fundamental strings lined up along one particular direction, and
turn on oscillators along each of those strings in such a way that
the oscillator energy density settles down to a uniform value
along each and every string. The proposed map is to identify
energy of electric flux lines as coming from the "winding" mode
part of fundamental strings, while attributing the tachyon matter
energy to the oscillator part. Static properties of the two sides
were considered and easily shown to be identical
\cite{Sen:2003xs}.

In this note, we will consider the two sides in detail and try to
develop more precise dictionary. Since we are trying to map a
system of two types of fluid, namely string fluid and tachyon
matter, to that of one kind of objects, namely classical closed
string, the map cannot be exactly one to one. Rather, there has to
be some ambiguity or redundancy in the translation. As we begin
detailed discussion below, this point will become clearer.

In section 2, we start with classical motion of a long Nambu-Goto
string and discuss stationary solutions and modulations of such
stationary solutions. Once a string is endowed with high frequency
oscillations along its length, nonlinear nature of Nambu-Goto
action sets in, which we will discuss in some detail. We focus on
energy-momentum flow along classical strings, and show that on
average the flow obeys a modified light-cone in a manner decided
by the averaged energy and momentum in the background oscillatory
strings. Section 3 will review properties of string fluid and
tachyon matter, and compare it against those of long Nambu-Goto
string(s). Again we will concentrate on small fluctuation around
stationary solutions, and find that the propagation of
disturbances mimics exactly its counterpart in Nambu-Goto strings
found in section 2. Section 4 discusses implications of our
finding. In section 5, we will make our argument more precise by
realizing tachyon matter as a mathematical device that encodes
energy and momentum of microscopic oscillations of a string. An
outlook is provided in section 6.

\section{Classical Motion of Long Nambu-Goto String}

We will consider an infinitely long string whose dynamics
is govern by the usual Nambu-Goto action. The action is
\begin{equation}
-\frac{1}{2\pi\alpha'} \int d\tau d\sigma\;\sqrt{-{\rm Det}\; g} ,
\end{equation}
with the induced metric $g$ on the world sheet,
\begin{equation}
g_{ij}=\partial_i X^\mu\partial_j X^\nu \eta_{\mu\nu} .
\end{equation}
Throughout this note we consider  flat spacetime only, hence
$\eta_{\mu\nu}={\rm diag}(-1,1,1,\dots,1)$, and furthermore
adopt the string unit, $2\pi\alpha'=1$.

Given that we are interested in long strings,
a convenient gauge choice for the worldvolume coordinate
is
\begin{equation}
\tau= X^0,\qquad \sigma= X^1 ,
\end{equation}
upon which we can write the induced metric as
\begin{equation}
g_{ij}=\eta_{ij}+\sum_{I=2}^{D-1}\partial_iX^I\partial_jX^I ,
\end{equation}
where $\partial X^I$ ($1< I\le D-1$) represents deviation
of the long string from the straight shape.

\subsection{Canonical Formulation}

Usual method of analyzing the motion of Nambu-Goto string
is to introduce intrinsic metric as a Lagrange multiplier
field and use the Polyakov action instead. While the
latter has obvious advantage in quantization procedure,
typical gauge choice there turns out rather inconvenient
for our purpose. We will stick to the Nambu-Goto form
and instead go to Hamiltonian formulation.
The Hamiltonian density is
\begin{equation}
H=\sqrt{1+\Pi_I\Pi_I+Y^I Y^I +(\Pi_I Y^I)^2}
\end{equation}
where sum over $I$ is implicit. $\Pi_I$ is the conjugate
momentum of $X_I$, while we introduced a shorthand notation
\begin{equation}
Y^I\equiv \partial_\sigma X^I  .
\end{equation}
Recall that the conserved Noether momentum density
on the worldsheet
is
\begin{equation}
P \equiv -\frac{\delta L}{\delta \dot{X}^I}\; \partial_\sigma X^I
=-\Pi_I\,Y^I .
\end{equation}
{}From canonical equation of motions, we find
\begin{eqnarray}
\dot \Pi_I=\partial_\sigma \left(\frac{Y_I -
P\Pi_I}{H}\right),\qquad \dot Y^I =
\partial_\sigma\left(\frac{\Pi^I - P Y^I}{H}\right) ,
\end{eqnarray}
while the worldsheet energy-momentum conservation may
be written as,
\begin{equation}
\dot H=-\partial_\sigma P,\qquad
-\dot P= \partial_\sigma\left(\frac{P^2-1}{H}\right) .
\label{emc}
\end{equation}

\subsection{Solutions}

We are primarily interested in generic perturbation of string
motion around some stable classical solutions. For this
let us find typical stationary solution first.
An important family of solutions for us are standing wave
solutions, i.e. those with
no Noether momentum density, $P=-\Pi_I Y_I=0$.
This, together with the energy-momentum conservation,
then implies
\begin{equation}
\dot H=0,\qquad
\partial_\sigma H=0 .
\end{equation}
We will denote this constant value of $H$ as ${\cal E}$
throughout this paper. The simplified equation of motion is
\begin{equation}
\dot \Pi_I=\frac{1}{\cal E}\,\partial_\sigma Y_I,\qquad
\dot Y^I = \frac{1}{\cal E}\,\partial_\sigma \Pi^I .
\label{static}
\end{equation}
While the equation may look linear superficially, it is
not the case because the constant energy density ${\cal E}
=(1+\Pi^2+Y^2)^{1/2}$
actually depends on specific solutions. The simplest solution
of this type may be found when we consider exciting two
transverse directions. For instance,
\begin{equation}
X^2+iX^3= \sqrt{\frac{{\cal E}^2-1}{{\cal E}^2\omega^2}}
\,\sin\left({\cal E}\omega\sigma\right)
e^{\pm i\omega\tau} ,
\label{solution1}
\end{equation}
describes a sinusoidal deviation from the straight string,
stabilized by centrifugal force of the rotation in the $X^{2,3}$
plane. This solves Eq.~(\ref{static}) and has $P=0$.

Another important, perhaps more familiar type of solutions involve
one-sided propagation of fluctuations. For this, rewrite the
Hamiltonian density as
\begin{equation}
H=\sqrt{\left(1\pm \Pi_IY^I\right)^2 +\left(\Pi_I\mp Y_I\right)^2} .
\label{bps}
\end{equation}
It is not difficult to see that solutions to the first order
equation
\begin{equation}
\Pi_I=\pm Y_I ,
\end{equation}
reduces the canonical equation of motion to
\begin{equation}
\dot \Pi_I=\pm \partial_\sigma \Pi_I ,\qquad
\dot Y_I=\pm \partial_\sigma Y_I ,
\end{equation}
which gives simple plane-wave solutions,
\begin{equation}
X^I={\rm Re}\left[C^I e^{i\omega(\sigma\mp \tau)}\right] ,
\end{equation}
with arbitrary complex numbers, $C^I$, either moving "up" or
"down" along the length of the string at unit speed, depending on
the sign. These plane-wave solutions are reminiscent of those used
as the basis for quantizing the string in the conformal gauge.
Note that our gauge choice is {\it not} conformal. Although the
nonlinearity of Nambu-Goto action is not manifest for these
solutions, this apparent linearity is the artifact of the ansatz.
The energy density and the momentum density of such solutions are
\begin{equation}
H\biggr\vert_{\Pi=\pm Y}=1+\Pi_I\Pi_I
,\qquad
P\biggr\vert_{\Pi=\pm Y}=\pm \Pi_I\Pi_I .
\end{equation}
We will call these solutions BPS type for an obvious reason.

\subsection{Modulating a Standing Wave}

As a concrete example of the phenomenon we will see repeatedly,
let us take our previous standing wave solution (\ref{solution1})
and perturb it. The aim of this exercise is two-fold. One is to
show that the propagation speed of a perturbation is generically
less than unit, perhaps contrary to naive expectation. The other
is to visualize two types of perturbations that we will identify
as motion of flux lines and also motion of tachyon matter,
respectively, in section 3.

For simplicity, we consider motion of string stretched along $X^1$
and oscillate on the $X^{2,3}$ plane only. Adopting the vector
notation $\vX=(X^2,X^3)$, the background standing wave solution
(\ref{solution1}) is
\be \vX_0=R(\sigma)\nhat(\tau)      , \ee
where
\be R(\sigma)=\sqrt{\frac{\CE^2-1}{\CE^2 \omega^2}}
\sin(\CE\omega\sigma)      ,\qquad \nhat(\tau)
=(\cos(\omega\tau),\sin(\omega\tau))      .
\ee
Throughout this
computation, we will consider the case where $\omega$ is large
while ${\cal E}$ remains finite. Note that in this limit the
amplitude of the standing wave scales as $1/\omega$ and can be
made arbitrarily small. Thus, when seen from far away, the string
will look like a thickened, straight string with mass density
larger than usual by the factor ${\cal E}$.

One can think of two types of modulations on this oscillating string,
"transverse" and "longitudinal," later to be identified with
motion of string fluid and motion of tachyon matter, respectively.
Of course the reason we can talk about "longitudinal" waves is
just that the string that moves about is no longer the bare
fundamental string. Rather it is an effective one involving
nontrivial solution of fundamental string which broke the
translation invariance along the fundamental string.

The former perturbation corresponds to introducing slow undulating
motion that involves bending in $X^{2,3}$ plane. If the
frequency/wavenumber of this modulation is small enough, the
energy involved would be also small even if the amplitude is
larger than than of the standing wave. From afar, the resulting
motion will appear as the thickened string of mass density $\cal
E$, stretched along $X^1$, oscillating transversally.  In the next
section, we will identify this with motion of flux lines.

\begin{figure}[t]
\begin{center}
\scalebox{0.8}[0.45]{\includegraphics{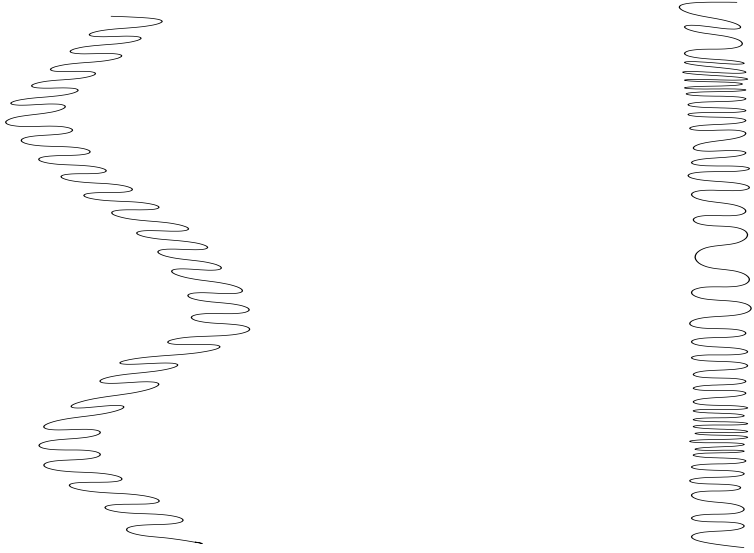}}
\par
\vskip-2.0cm{}
\end{center}
\begin{quote}
\caption{\small Pictorial representations of two typical perturbations
are drawn here. The unperturbed solution would corresponds to high
frequency configurations with no modulation. The left side represents
"transverse" mode where the fast oscillating string experiences
a large scale undulation. The right hand side represents "longitudinal"
mode where the high frequency background has its frequency modulated
along the length of the string. }
\end{quote} \label{string}
\end{figure}
Let us consider such "transverse" perturbation first. For this,
write  the perturbed solution as $\vX=\vX_0+\dvX$ and expand the
Nambu-Goto Lagrangian up to second order in $\dvX$. For
sufficiently large-scale fluctuations $\dvX$, we will perform
suitable space-time average with respect to the standing wave
background to get the effective Lagrangian for $\dvX$. The
Nambu-Goto Lagrangian is
\be \CL=-\sqrt{\Big(1-(\p_\tau
\vX)^2\Big)\Big(1+(\p_\sigma \vX)^2\Big)+ \Big(\p_\tau
\vX\cdot\p_\sigma\vX\Big)^2}      , \ee and the second order
expansion looks like \bear \CL&\approx&
-\CE(1-\omega^2R^2)+\frac{\CE}{2}
\dXtau^2-\frac{1}{2\CE}\dXprime^2\nonumber\\
&&-\frac{1}{\CE(1-\omega^2R^2)}
\Big(-2\omega R R'(\thetahat\cdot\dXtau)(\nhat\cdot\dXprime)+
\frac{1}{2}(\omega R \thetahat\cdot \dXprime+R'\nhat\cdot\dXtau)^2\Big)
\nonumber\\
&&+\frac{\CE\omega^2R^2}{2(1-\omega^2R^2)}(\thetahat\cdot\dXtau)^2+
\frac{(R')^2}{2\CE^3(1-\omega^2R^2)}(\nhat\cdot\dXprime)^2+ \ldots
, \eear where $\thetahat=(-\sin(\omega\tau),\cos(\omega\tau))$,
$R'=\p_\sigma R$, $\dXtau=\p_\tau\dvX$, and
$\dXprime=\p_\sigma\dvX$. Performing space-time average, for
example, \be
\langle\nhat^i\nhat^j\rangle=\langle\thetahat^i\thetahat^j\rangle
=\frac{1}{2}\del^{ij}       ,\quad \bigg\langle \frac{R
R'\nhat^i\thetahat^j}{1-\omega^2R^2} \bigg\rangle=0      , \ee we
have the effective action for $\dvX$, \be \CL\approx
-\frac{1+\CE^2}{2\CE}+C \cdot\Big((\dXtau)^2
-\frac{1}{\CE^2}(\dXprime)^2\Big)+\cdots , \ee where the constant
$C$ is
\be C=\frac{\CE}{2}\bigg(1-\frac{1}{2\CE^2}\Big\langle
\frac{-1+\CE^2-2\CE^2\omega^2R^2}{1-\omega^2R^2}\Big\rangle\bigg)
. \ee The above effective action shows  that the propagation
velocity is reduced to ${1}/{\CE}$.

We now turn to the "longitudinal" modulation case. Imagine that we
slowly change the frequency energy density of the standing wave
along $X^1$ direction. At any given time, the configuration still
looks like a thickened string but whose mass density varies along
$X^1$. Of course the variation of energy density will propagate up
and down the thickened string, and behaves like an longitudinal
wave. For this reason, we will call this second type
"longitudinal" perturbation. In next section, we will identify
this with motion of tachyon matter.

Such "longitudinal" mode can be realized in many different ways.
We choose the following form; \be
\vX=\sqrt{\frac{\CE^2-1}{\CE^2\omega^2}}\sin\Big(\CE\omega\sigma
+f(\tau,\sigma)\Big)\,\, \nhat(\tau)      , \ee where the
modulation mode $f$ is slowly varying. Taking space-time
derivatives gives \bear \vX_\tau&=&\omega R
\,\,\thetahat+f_\tau\frac{R'}{\CE\omega}
\,\,\nhat,\nonumber\\
\vX'&=&\Big(1+f'\frac{1}{\CE\omega}\Big)R' \,\,\nhat      , \eear
where $R(\sigma)$ is as before. More precisely, the argument in
$R$ is modified by $f$, but for sufficiently slowly varying $f$,
we can absorb it by a shift in $\sigma$ for a sizable region of
space-time where we are going to average. Its presence thus makes
no change in our final averaged effective action. Hence, for our
purpose of calculating the effective action for $f$, it makes no
difference to use
\bear
\vX_\tau&=&(\vX_{0})_\tau+f_\tau\frac{R'}{\CE\omega}
\,\,\nhat,\nonumber\\
\vX'&=&\vX'_0+f'\frac{R'}{\CE\omega} \,\,\nhat      . \eear After
a straightforward calculation, the second order expansion reads as
\be \CL\approx -\CE(1-\omega^2R^2)+
\frac{-1+\CE^2-\CE^2\omega^2R^2}{2\CE^3\omega^2(1-\omega^2R^2)}
\bigg((f_\tau)^2-\frac{1}{\CE^2}(f')^2\bigg)+\cdots      , \ee
whose space-time average is \be \CL\approx
-\frac{1+\CE^2}{2\CE}+C'\cdot\bigg
((f_\tau)^2-\frac{1}{\CE^2}(f')^2\bigg)+\cdots      . \ee This
again reproduces the reduced speed of ${1}/{\CE}$.

\subsection{Perturbing Energy and Momentum }

We want to ask the following question: Take a collection of highly
oscillatory strings which are roughly straight and collinear, and
perturb them by introducing a common low frequency modulation. How
does such perturbation propagate along the length of the strings?
In the last subsection, we considered this question for two very
specific classes of stationary Nambu-Goto string solution and
found the answer
\begin{equation}
\hbox{effective speed of light} =\frac{1}{\cal E} \quad ,
\end{equation}
which is the ratio between energy density of a straight string and
that of stretched oscillatory string, per unit distance in
spacetime along the string. In this subsection, we will show that
this property is not particular to the solutions used, but is much
more generic if we consider a large number of nearly coincident
strings.

For this, let us concentrate on basic physical properties
such as how energy and momentum of the perturbing mode
flows.  The above result
for the standing wave can be understood in a simple and
universal manner in this way. For the standing wave solution
($P=0$), the first order perturbation of energy momentum
conservation (\ref{emc}) gives,
\begin{equation}
\delta \dot H =-\partial_\sigma \delta P,\qquad
-\delta \dot P =\frac{1}{{\cal E}^2}\,\partial_\sigma \delta H .
\end{equation}
Therefore, regardless of details of the perturbation solution, the
perturbed energy and momentum obeys a sort of an 1+1 dimensional
Klein-Gordon equation;
\begin{equation}
\left[{\cal E}^2\partial_\tau^2-\partial_\sigma^2\right]
\delta H=0,\qquad
\left[{\cal E}^2\partial_\tau^2-\partial_\sigma^2\right]
\delta P=0 ,
\end{equation}
whose propagation velocity is $\pm 1/{\cal E}$, inverse of energy
density of the standing wave type solution. Remembering the
gauge choice for the worldsheet diffeomorphism, $\tau=X^0$
and $\sigma=X^1$, this means that any modulation of the
highly oscillatory long string will propagate at reduced
speed of light along $X^1$ direction, as seen by any spacetime
observers. This reduced "speed of light" can be rewritten
more suggestively as,
\begin{equation}
\frac{1}{\cal E}\;=\; \frac{\hbox{String Charge per Unit
 $X^1$ Length}}{\hbox{Energy per Unit $X^1$ Length}} \:\le \:1 ,
\end{equation}
which will be useful later.

Heuristically this reduction may be understood from the fact
that per unit $X^1$ length, the actual length of string is
elongated due to the high frequency oscillation. Propagation
speed would be unit on the worldsheet when measured with
respect to actual induced metric. However, the propagation
speed we are discussing above is measured by $dX^1/dX^0$
which is different from the worldsheet speed.

For more general class of oscillatory solutions, the above
derivation will not be applicable directly because the background
energy-momentum is not itself constant. On the other hand, if we
are primarily interested in slow modulation of energy and
momentum, we may as well "average" the background quantities to
extract low frequency part only. Then, what do we mean by
averaging? One way is to average over time scale $\sim 1/w$ where
$w$ is typical high frequency of the oscillatory string. While
this might be the simplest conceptually, the averaging procedure
we will adopt is actually hinted by the unstable D-brane physics.

Here we are attempting to match the classical low energy system of
a unstable D-brane to classical Nambu-Goto strings. In the former
system, the initial energy content scales as $1/g_s$, which is
arbitrarily large in classical limit. This means that the final
state must be composed of a larger number of closed string degrees
of freedom, if there is such an interpretation at all. Thus
instead of averaging over time along a single string, we may
consider a bundle of fundamental strings and average over the
nearby closed strings. In this spirit, Let us replace
\begin{eqnarray}
H+\delta H&\rightarrow& {\cal E} +\delta {\cal H}, \nonumber \\
P+\delta P&\rightarrow& {\cal P}+\delta {\cal P},
\end{eqnarray}
where the calligraphic characters denote averaged or low frequency
part only. More precise definition of what we mean by this
averaging will be discussed in section 5. We will consider only
the case where ${\cal E}$ and ${\cal P}$ are constants in this
approximate sense. The BPS plane-wave solution of previous
subsection with large $\omega$ would be one such example.

The modulation
is encoded in $\delta \cal H $ and $\delta \cal P$ which we must assume
to have only slow dependence
on $\tau $ and $\sigma$. The latter quantities then obey
the averaged conservation equation,
\begin{equation}
\delta \dot {\cal H}=-\partial_\sigma \delta {\cal P},\qquad
-\delta \dot {\cal P}=2\,\frac{\cal P}{\cal E}\;\partial_\sigma\delta {\cal P}\;+\;
\left(\frac{1-{\cal P}^2}{{\cal E}^2}\right)\partial_\sigma\delta {\cal H} .
\end{equation}
Second order equation for $\delta {\cal P}$ is then
\begin{equation}
\left[{\cal E}^2\partial_\tau^2 + 2{\cal EP}
\partial_\tau\partial_\sigma-(1-{\cal P}^2)
\partial_\sigma^2\right]\delta {\cal P}=0 .
\end{equation}
Extracting the dispersion relation from this, we can solve
for relationship between low frequency $\Omega$ and low
wavenumber $K$.
\begin{equation}
\delta {\cal H},\delta {\cal P} \sim e^{-i\Omega \tau +i K\sigma} .
\end{equation}
The result is
\begin{equation}
\frac{\Omega}{K}=\frac{\cal P}{\cal E}\pm \frac{1}{\cal E} .
\label{stringdispersion}
\end{equation}
Again the effective "speed of light" as seen by the spacetime
observer deviate from unit. Furthermore, modulation speed is
asymmetrical; disturbances tends to move faster along the
direction of the background momentum $\cal P$ than along the
backward direction. A simple consistency check is to show that the
faster of the two speeds does not exceed unit, set by speed of
light, 1.  This can be see easily using the general formula
(\ref{bps}),
\begin{equation}
\frac{|\CP \pm 1|}{\cal E}
=\frac{|\langle P\rangle \pm 1|}{\langle H\rangle}
=\frac{|1\pm \langle P\rangle |}{\left\langle
\sqrt{(1\pm P)^2 +(\Pi_I\pm Y_I)^2}\right\rangle}
\:\le\:\frac{|1\pm \langle P\rangle |}{\left\langle
\sqrt{(1\pm P)^2}\right\rangle} =1 .
\end{equation}
In next section , we will show that
this causal properties are exactly matched by the combined
system of string fluid and tachyon matter.

\section{String Fluid and Tachyon Matter}

Tachyon condensation of an unstable D-brane with conserved
electric flux on it has been analyzed extensively using an
effective field theory of gauge field coupled to tachyon
\cite{Gibbons:2000hf,Gibbons:2002tv,Kwon:2003qn,Sen:2000kd}.\footnote{A
fluid-like behavior of Born-Infeld action in the string coupling
limit, i.e., in the limit of vanishing tension, was previously
observed in
Ref.~\cite{Lindstrom:1997uj,Gustafsson:1998ej,Lindstrom:1999tk}.}As
the tachyon potential approaches to zero, the most convenient
description of the system turns out to be canonical Hamiltonian
formalism. In Ref.~\cite{Kwon:2003qn}, all static with no net
momentum solutions have been characterized as straight conserved
flux lines with tachyon matter distributed uniformly along the
flux direction. Transverse variations of both flux density and
tachyon matter are unconstrained, implying huge degeneracy of the
solutions.

In this section, we perform similar analysis to the previous section
of looking at small perturbation spectrum
for stationary states of string fluid with tachyon matter.
As in the Nambu-Goto string case, we
include the situation where the system has a constant
momentum density whose direction is along the flux line.
The existence of this type of configuration is obvious by boosting
the solution with no momentum along
the string direction. In fact, the dispersion relation of
perturbations for this system may be obtained
by simple Lorentz transformation. We will check the
consistency our results with Lorentz symmetry.

We will then compare the results of this section with those in the
previous section for Nambu-Goto string, and find intriguing
similarity between them. Pushing forward this observation, we
explore a possibility of identifying the system of flux lines to a
macroscopic collection of fundamental strings, with the tachyon
matter  interpreted as an effective degree of freedom describing
small fast oscillations of fundamental strings.

\begin{figure}[t]
\begin{center}
\scalebox{0.7}[0.55]{\includegraphics{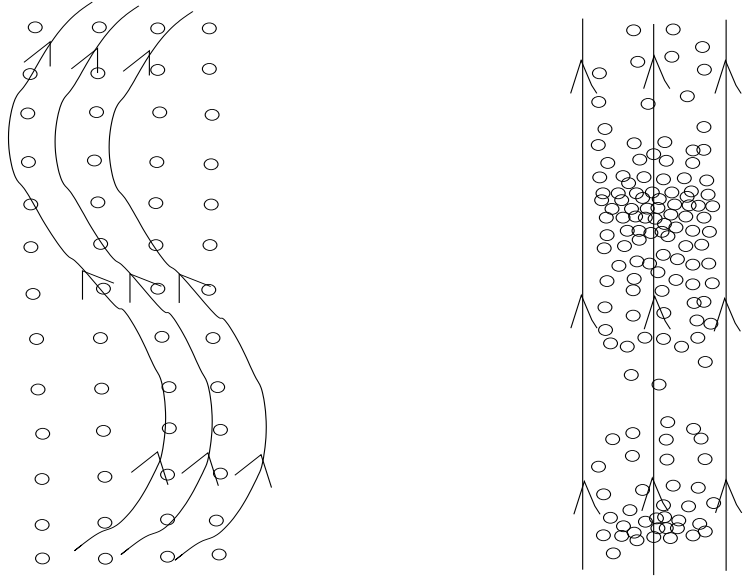}}
\par
\vskip-2.0cm{}
\end{center}
\begin{quote}
\caption{\small  Perturbation associated with the two types of fluids are
drawn here. The small circles represent tachyon matter, while the lines
are string fluid. In the background of static solution, two allowed
perturbations are transverse motions of string fluid itself (left) and
arbitrary motion of tachyon matter (right). The two map to closed
string picture, respectively,
as "transverse" and "longitudinal" modulation of collection
of fundamental strings with random high frequency modes turned on.   }
\end{quote} \label{fluid}
\end{figure}

\subsection{Stationary Solutions}

We start from the effective action for an unstable Dp-brane,
including tachyon as a dynamical degree of freedom \be -\int
d^{p+1}x\,\, V(T)\sqrt{-\det(\eta_{\mu\nu} +F_\mn+\p_\mu T\p_\nu
T)}      . \ee Upon tachyon condensation $V\ra 0$, the system is
described in canonical formalism by the Hamiltonian
\cite{Gibbons:2000hf} \be \CH=\sqrt{\pi^i\pi^i+\pi^T \pi^T+(\p_i
T\pi^i)^2+\CP_i\CP_i}   \quad    , \label{hamil} \ee where
$\pi^i,\,\pi^T$ are conjugate momenta of gauge fields and tachyon,
and $\CP_i=-F_{ij}\pi^j-\p_i T\pi^T$ are the momentum densities.
For future convenience, let us recapitulate the energy-momentum
tensor of the theory in tachyon condensation limit
\cite{Gibbons:2002tv}, \bear
T_{00}&=&\CH      ,\nonumber \\
T_{0i}&=&-\CP_i      ,\nonumber\\
T_{ij}&=&\frac{-\pi^i\pi^j+\CP_i\CP_j}{\CH}      . \eear

It is rather straightforward to derive the equations
of motion from the above Hamiltonian. However,
there is a more intuitive description of the system
in terms of fluid-like variables \cite{Gibbons:2000hf},
whose existence
reflects an intrinsic property of the system as some
kind of string fluid.
Define ( Indices $i,j,k$ run from $1$ to $p$)
\bear
n^k=\frac{\pi^k}{\CH}\quad&,&\qquad n^T=\frac{\pi^T}{\CH},\nonumber\\
v^k=\frac{\CP_k}{\CH}\quad&,&\qquad v^T=\frac{\p_iT \pi^i}{\CH},
\eear
Then, the half of canonical equations of motion combined
with energy-momentum conservation
gives
\bear
\p_0 n^k+v^i\p_in^k&=&n^i\p_i v^k      ,\nonumber\\
\p_0n^T+v^i\p_in^T&=&n^i\p_i v^T      ,\nonumber\\
\p_0 v^k+v^i\p_iv^k&=&n^i\p_in^k      ,\nonumber\\
\p_0v^T+v^i\p_iv^T&=&n^i\p_in^T      , \label{fluideq} \eear
together with constraints, \be n^in^i+n^Tn^T+v^iv^i+v^Tv^T=1
,\qquad n^iv^i+n^Tv^T=0      , \label{constraint} \ee and an
integrability condition that $F_{ij}$ and $\p_i T$, which can be
solved from the other half of canonical equations of motion with
the above fluid background, should satisfy \bear
v^i&=&-F_{ij}n^j-n^T\p_iT      ,\nonumber\\
v^T&=&\p_iT n^i      . \eear For the case of static ($\p_0=0$,
$\CP_i=0$) configurations, it is not difficult to find the general
form of solutions \cite{Kwon:2003qn}. The flux density with
specified constant direction, say $\pi^1$ (or $n^1$), and the
tachyon matter $\pi^T$ (or $n^T$) are the only non-vanishing
quantities in this case. They may have arbitrary transverse
profiles to $x^1$, but restrained to be constant along $x^1$.
Complete characterization of these solutions up to overall
rotation is given by the following three nontrivial quantities,
\begin{eqnarray}
\CH=\CH(x^2,\dots,x^p),&&\nonumber \\
n^1=n^1(x^2,\dots,x^p),&& n^T=n^T(x^2,\dots,x^p),
\end{eqnarray}
while all others vanish.
Magnetic fields $F_{ij}$ and tachyon gradient $\p_iT$
are absent in these solutions.

We are interested in slightly more general situation than this,
however. We have in mind configurations in which there is a net
momentum density along $x^1$, whose existence may easily be
inferred by boosting the static case. Once we turn on $\CP_1$ (or
$v^1$), Eq.(\ref{constraint}) tells us that we need to turn on
$v^T$ simultaneously, indicating a constant tachyon gradient
$\p_1T$. Because we boosted the system without magnetic fields
along the direction of electric field, we still do not have any
magnetic fields, $F_{ij}=0$, henceforth gauge fields do not
contribute to the momentum density; \be
\CP_1=-F_{1j}\pi^j-\p_1T\pi^T=-\p_1T \pi^T       . \ee The
Hamiltonian (\ref{hamil}) becomes \bear
\CH&=&\sqrt{(\pi^1)^2+(\pi^T)^2+(\p_1T\pi^1)^2+(\p_1T\pi^T)^2}\nonumber\\
&=&\sqrt{(\pi^1)^2+(\pi^T)^2}\cdot \sqrt{1+(\p_1T)^2}      , \eear
and the constraints (\ref{constraint}) are easily satisfied. Thus
most general stationary solution is characterized by
\begin{eqnarray}
\CH=\CH(x^2,\dots,x^p),&&\nonumber \\
n^1=n^1(x^2,\dots,x^p),&&n^T=n^T(x^2,\dots,x^p) ,\nonumber \\
v^1=v^1(x^2,\dots,x^p),&&v^T=-n^1 v^1/n^T.
\end{eqnarray}
This latter class of solution cannot be extended to everywhere
because the gradient of $T$ will eventually introduce a region
where $V\neq 0$, and possibly domain walls, leading us out of the
region of validity for the above equation of motions. However,
we are mainly interested in general motion of string fluid
in the vacuum, and local solutions are enough.

\subsection{Perturbation}

A linear perturbation to the fluid equation (\ref{fluideq})
in the background of the previous
subsection reads
\bear
\p_0\del n^\alpha+v^1\p_1\del n^\alpha
-n^1\p_1 \del v^\alpha&=&0      ,\\
\p_0\del v^\alpha+v^1\p_1 \del v^\alpha -n^1\p_1 \del n^\alpha&=&0
      , \eear where $\alpha$ runs not only from $1$ to $p$, but
also the tachyon direction $T$. Remember that the background
quantities, like $n^1$ and $v^1$, are constant along $x^1$
direction. Taking time derivative of the first equation and
replacing $\del v^\alpha$ in favor of $\del n^\alpha$ by judicious
use of the two equations produce \be \p_0^2\del
n^\alpha+2v^1\p_0\p_1\del n^\alpha +((v^1)^2-(n^1)^2)\p_1^2\del
n^\alpha=0      , \ee and the same equation for $\del v^\alpha$.
The resulting dispersion relation between frequency $\Omega$ and
wave number $K$
\begin{equation}
\delta n, \delta v \sim e^{-i\Omega t +iKx^1},
\end{equation}
is \be \frac{\Omega}{K}\,\,=\,\,v^1\pm n^1\,\,
=\,\,\frac{\CP_1}{\CH}\,\pm\,\frac{\pi^1}{\CH}      .
\label{fluiddispersion} \ee
We now notice a striking similarity of the
above dispersion relation
to that of Nambu-Goto strings, Eq.(\ref{stringdispersion}).
In fact, a bundle of flux lines
with unit string charge has a line-energy density
$\CE={\CH}/{\pi^1}$ and a line-momentum density
$\CP={\CP_1}/{\pi^1}$, in terms of which the two
formulae match precisely. In the next section,
we further elaborate on possible relations between
Nambu-Goto strings and string fluid with tachyon matter.

It is not particularly obvious how such a formula for
velocity is consistent with laws of relativity. As in
the case of fundamental string, we can ask whether
the modified propagation speed never exceeds unit.
This  follows from the constraint on
$n$ and $v$
\begin{equation}
0 \le ( v^1 \pm n^1)^2
= (1-n_T^2+v_T^2) \mp n_T v_T=1-(n_T\pm v_T)^2 \le 1 .
\end{equation}
As a further consistency check let us show that the dispersion
relation, (\ref{fluiddispersion}), transforms under Lorentz boost
as it should. $\Omega/K$ represents the velocity of small
fluctuation signals, and we know how velocities should add under
boost transformations. Under a Lorentz boost along direction $x^1$
with velocity $\beta$, a velocity $u$ along $x^1$ transform as \be
u'=\frac{u-\beta}{1-\beta \,u}      . \ee Suppose a system with
constant $\CH$  and $\CP_1$, in which perturbation signals are
claimed to propagate with the velocity
$u=\frac{\CP_1\pm\pi^1}{\CH}$. Now consider another reference
frame which is moving along $x^1$ direction with a velocity
$\beta$. Recalling that $T_{11}=\frac{-(\pi^1)^2+(\CP_1)^2}{\CH}$
is present in the system, the energy and momentum density of the
moving system are given by \bear
\CH'&=&\gamma^2T_{00}+2\gamma^2\beta T_{01}
+\gamma^2\beta^2 T_{11}\nonumber\\
&=&\gamma^2\CH-2\gamma^2\beta\CP_1
+\gamma^2\beta^2\frac{-(\pi^1)^2+(\CP_1)^2}{\CH}     \quad ,\\
\CP'_1&=&-\gamma^2\beta T_{00}
-\gamma^2(1+\beta^2)T_{01}-\gamma^2\beta T_{11}\nonumber\\
&=&-\gamma^2\beta\CH+\gamma^2(1+\beta^2)\CP_1 -\gamma^2\beta
\frac{-(\pi^1)^2+(\CP_1)^2}{\CH}    \quad  , \eear where
$\gamma^2=1/(1-\beta^2)$. For this particular class of Lorentz
boost, it turns out that $\pi'^1=\pi^1$. With a little bit of
algebra, then, it is straightforward to show that \be
u'\,\,=\,\,\frac{\CP'_1}{\CH'}\,\pm\, \frac{\pi'^1}{\CH'}\,\,
=\,\,\frac{\Big(\frac{\CP_1}{\CH}\pm\frac{\pi^1}{\CH}\Big)-\beta}
{1-\beta\Big(\frac{\CP_1}{\CH}\pm\frac{\pi^1}{\CH}\Big)}\,\,
=\,\,\frac{u-\beta}{1-\beta \,u}   \quad   . \ee just as a velocity
should.

\section{Closed Strings Picture: Classical Limit}

In section 2, we have seen that small fluctuations on top of
oscillatory Nambu-Goto strings travel with the universal velocity,
$u={\CP}/{\CE}\pm{1}/{\CE}$, without any regard to the character
of perturbations, as long as they are slowly varying with large
wavelengths compared to the background oscillation. An observer
with her low-energy, large-scale eyes may be able to see
large-scale transverse fluctuations, but wouldn't have ability to
tell microscopic details of oscillatory nature of the string,
except its energy and momentum. Hence, a large-scale observer may
describe the system in terms of transverse string distortions plus
some new effective degree of freedom that carries energy-momentum
of averaged microscopic modes.

\begin{figure}[t]
\begin{center}
\scalebox{0.25}[0.4]{\includegraphics{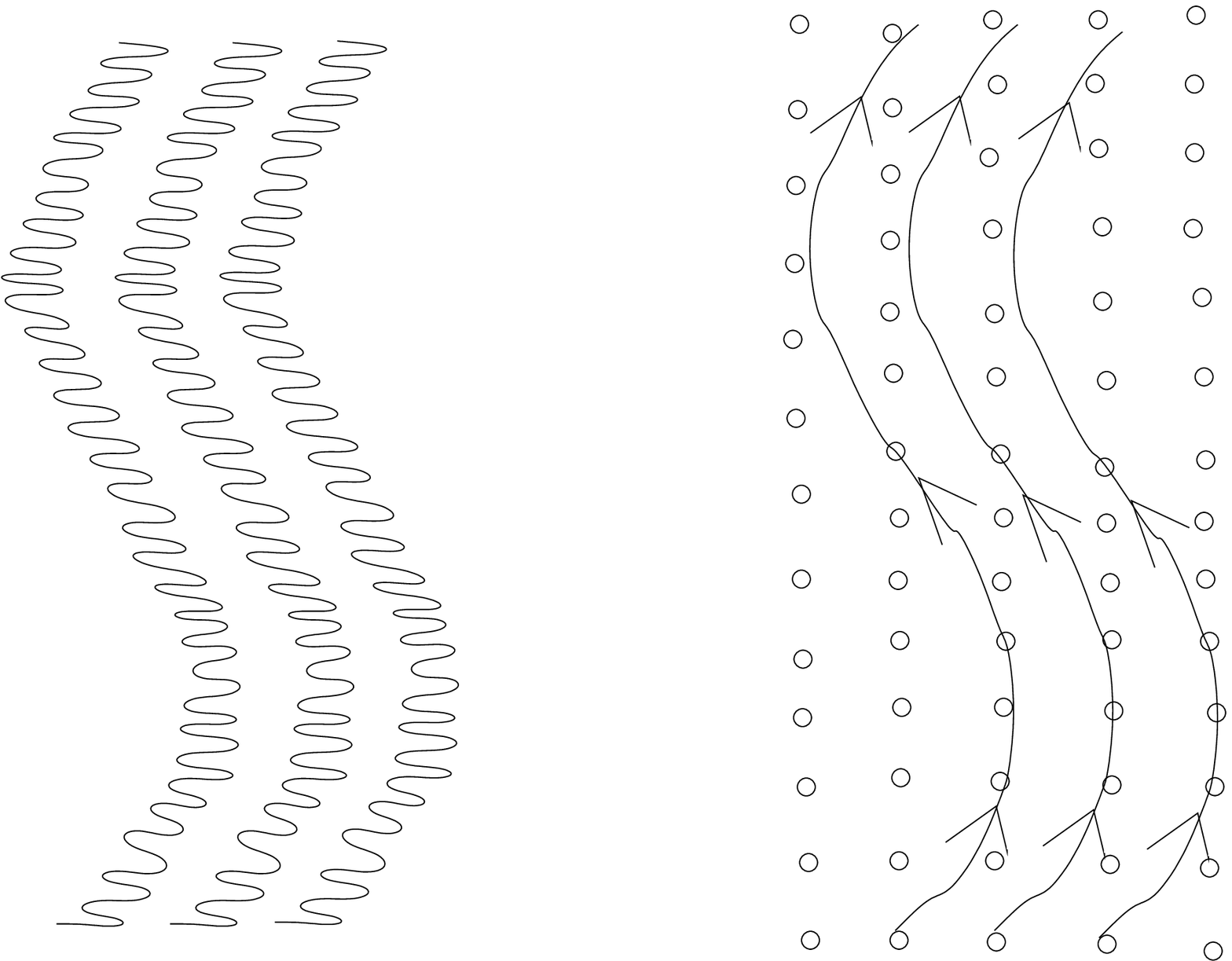}
\qquad\qquad\qquad\qquad\qquad\qquad\qquad\qquad\qquad
\qquad\qquad\qquad\qquad\qquad\qquad\qquad
\includegraphics{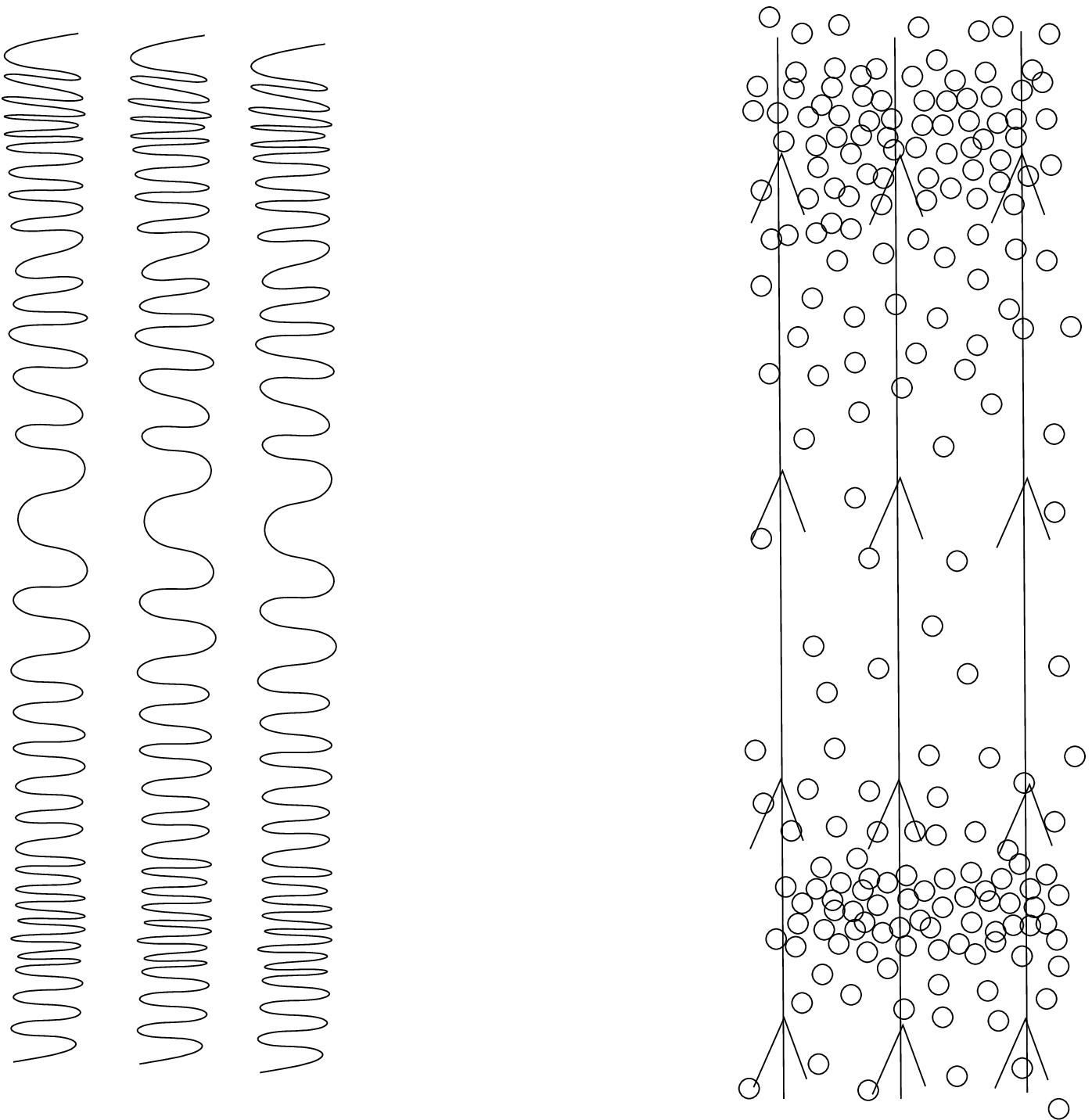}}
\par
\vskip-2.0cm{}
\end{center}
\begin{quote}
\caption{\small Identifications between modes of Nambu-Goto
strings (lefthand side in each pair) and fluctuations of electric
flux lines with tachyon matter (righthand side in each pair).
Lines with arrows represent electric flux while circles represent
tachyon matter.}
\end{quote} %\label{}
\end{figure}
On the other hand, analysis in  section 3 revealed a compelling
interpretation of flux lines coupled to tachyon matter as a
macroscopic collection of oscillatory strings. Perturbations of
the system consist of transverse distortions of flux lines plus
additional mode of tachyon matter fluctuation. Their propagation
velocity is again universal, and in fact, it is identical to that
of Nambu-Goto string when expressed in terms of properly
normalized energy-momentum line-density for unit string charge. We
are thus naturally led to relate tachyon matter perturbation in
string fluid side with the effective scalar degree of freedom on
Nambu-Goto string that encodes microscopic longitudinal
fluctuations. This complete matching of classical stationary
states and perturbation thereof indicates that this additional
degree of freedom is effectively represented by the tachyon matter
on the open string side.

With this, it is obvious that the macroscopic interpretation of
the string fluid/tachyon matter hybrid is consistent at least at
the level of classical limit. What we have  gathered here are
compelling evidences that the combined system of string fluid and
tachyon matter in the open string side actually describes
classical motion of stretched fundamental strings.

The interpretation of string fluid as macroscopic snapshot of
collection of many fundamental string  was quite natural from its
classical dynamics, and in fact was already suggested in
Ref.~\cite{Gibbons:2000hf}. On the other hand, interpretation of
tachyon matter is less straightforward. In this note, we are
advocating the viewpoint taken in Ref.~\cite{Sen:2003xs}, where
the energy associated with tachyon matter is attributed to high
frequency oscillations of the stretched fundamental strings
themselves.

This identification was suggested by first considering static and
stationary states of two sides \cite{Sen:2003xs}, and here we
strengthened the match by further comparing the behavior of small
fluctuations. The motion of string fluid is mapped to large scale
motion of stretched strings, while motion of tachyon matter is
shown to encode energy-momentum content of high frequency
oscillations on the same strings. The system of string fluid and
tachyon matter is then interpreted as a coarse-grained view of
macroscopic number of stretched fundamental strings. We find these
identifications between open string side and closed string side
quite compelling.

\section{Tachyon Matter from Coarse Graining of Fundamental Strings}

In the previous sections, the matching of fluctuation dynamics
between a macroscopic collection of Nambu-Goto strings and the
system of string fluid with tachyon matter was, in itself, a
compelling evidence of our claim. In this section, we try to make
a further step illustrating how tachyon matter degrees of freedom
may arise on the Nambu-Goto side via some coarse graining
procedure.

While this should be possible in general setting, we will
illustrate it by matching variables with the case of unstable D1
decay. Start with the unstable D1 action
\begin{equation}
-\int dx^2\; V(T)\sqrt{-{\rm Det}\left(\eta_{jk}+\p_j Z^I\p_k Z^I+ F_{jk}
+\p_j T\p_k T\right)} \quad ,
\end{equation}
with transverse scalars $Z^I$. Because D1 is also 1+1 dimensional
object, it is sensible to use the same index $I=2,3,\dots,D-1$ as
in Nambu-Goto action of fundamental string.

Let us put ${\cal N}$ unit  of fundamental string
charge, that is, electric flux $\pi_1={\cal N}$ on the
unstable D1, and let the combined system decay. The final
form of Hamiltonian is
\begin{equation}
\CH=\sqrt{{\cal N}^2+ \pi_I^2+\pi_T^2+ {\CP}_1^2+{\CP_I}^2+{\CP}_T^2} \quad ,
\label{fluidhamiltonian}
\end{equation}
where
$\pi_I$ and $\pi_T$ are conjugate momenta of transverse scalars
$Z^I$ and tachyon $T$, respectively. The other three quantities are
\begin{eqnarray}
\CP_1&=& -\partial_1 Z^I\pi_I-\partial_1 T\pi_T ,\nonumber \\
\CP_I&=& {\cal N}\partial_1 Z^I ,\nonumber\\
\CP_T&=& {\cal N}\partial_1 T .
\label{fluidmomenta}
\end{eqnarray}
Equations of motions for this system look simpler when we again
introduce fluid variables,
\begin{eqnarray}
(n_s)=\left(\frac{\cal N}{\CH}, \frac{\pi_I}{\CH}, \frac{\pi_T}{\CH}\right),
\qquad
(v_s)=\left(\frac{\CP_1}{\CH}, \frac{\CP_I}{\CH}, \frac{\CP_T}{\CH}\right),
\end{eqnarray}
with $s$ running over $1, I=2,\dots,D-1$ and $T$. The complete
low energy dynamics is encoded in
\begin{eqnarray}
{\dot n}_s+v^1\partial_1 n_s= n^1\partial_1 v_s,\qquad {\dot
v}_s+v^1\partial_1 v_s= n^1\partial_1 n_s . \label{fluidequation}
\end{eqnarray}
Note that $s=1$ part of these equations is nothing but the
energy-momentum conservation, once we note that Gauss law is
readily incorporated in $\pi_1={\cal N}$.

In the weak coupling limit, $g_s \rightarrow 0$, the initial
amount of energy on unstable D1-brane is large $\sim 1/g_s$. This
energy must be shared by $\pi_1$, $\pi_T$, etc, which means that
typical value of ${\cal N}\sim 1/g_s$ is also large. On the other
hand, ${\cal N}$ is the fundamental string charge and the
corresponding closed string picture should involve many nearly
aligned fundamental string moving under Nambu-Goto dynamics. We
wish to recover this form of low energy dynamics, in particular
the part associated with tachyon matter, by coarse-graining the
Nambu-Goto dynamics of many long strings.

To compare with the motion of fundamental string, consider ${\cal
N}$ infinitely stretched Nambu-Goto strings, whose Hamiltonian is
the simple sum of ${\cal N}$ separate Nambu-Goto strings,
\begin{equation}
H_{\cal N}=\sum_{a=1}^{\cal N} H^{(a)}.
\end{equation}
We labelled the strings by the index $(a)$. Interestingly, each
Nambu-Goto dynamics admits variables which behave similarly as the
fluid variables in the string fluid system, \be N^1\equiv
\frac{1}{H}\,\,,\,\,\,\,N^I\equiv
\frac{Y^I}{H}\,\,,\,\,\,\,V^1\equiv\frac{P}{H}
\,\,,\,\,\,\,V^I\equiv\frac{\Pi_I}{H} \;     , \ee where the
superscript in $N^1$ and $V^1$ means world-sheet $\sigma$
direction, while we suppressed $(a)$ index since every single
string obeys the same equation of motion, decoupled from each
other: They satisfy ( $m=(1,I)$ ) \bear
\p_\tau N^m+V^1\p_\sigma N^m&=&N^1\p_\sigma  V^m      ,\nonumber\\
\p_\tau V^m+V^1\p_\sigma V^m&=&N^1\p_\sigma N^m      .
\label{NGfluid} \eear By definition, they are constrained by \bear
(N^1)^2+(N^I)^2+(V^1)^2+(V^I)^2&=&1      ,\nonumber\\
N^1V^1+N^IV^I&=&0      . \label{NGconstraint} \eear Because we can
solve $N^1$ and $V^1$ from the above constraints, the independent
degrees of freedom are just transverse fluctuations, as should be
obvious from the original Hamiltonian.

We now consider the situation where these ${\cal N} (\gg 1)$
fundamental strings are roughly coincident, and try to derive a
coarse-grained system of equations which involve macroscopic
motion of the bundle of strings. High frequency oscillations of
individual strings are not something we can see in the coarse
grained dynamics, and thus will be deemed to be completely random.
For instance, we could imagine a collection of BPS type solutions
of section 2 with arbitrary initial conditions and arbitrary high
frequency $\omega$ for each string. In such a collection, half of
strings will have $\omega \gg 1$ and the other half, $-\omega \gg
1$. In contrast, we are interested in overall slow motion of the
bundle, so low frequency motion of strings are taken to be
identical.

To illustrate the averaging procedure, let us take the equation
(\ref{NGfluid}) and express it in a Fourier transformed basis by
swapping $\sigma$ in favor of its conjugate variable,
\begin{equation}
N^m=\frac{1}{2\pi}\int dp \;\hat N^m(p;\tau)\,e^{ip\sigma},\qquad
V^m=\frac{1}{2\pi}\int dp \;\hat V^m(p;\tau)\,e^{ip\sigma},
\end{equation}
we find equations like
\begin{equation}
\partial_\tau {\hat N}^m(k)+i\int  \left(\hat V^1(k-p)\hat N^m(p)
-\hat N^1(k-p) \hat V^m(p)\right)\,p\; dp=0      .
\end{equation}
We will consider only part of this equation and restrict our
attention to small $k$. Imagine that the motion consists of two
separate components; one with small $p\sim k$ corresponding to
slow modulations under which ${\cal N}$ strings move together and
the other with very large $p\rightarrow \omega$ with $|\omega| \gg
|k|$ under which each string moves independently. Then we may
split the integral in two parts, according to the range of
$p$-integration,
\begin{eqnarray}
&&\partial_\tau {\hat N}^m(k)+i\int_{small}  \left(\hat V^1(k-p)\hat N^m(p)
-\hat N^1(k-p) \hat V^m(p)\right)\,p\; dp \nonumber \\ \nonumber \\
&&=
-i\int_{large}  \left(\hat V^1(k-\omega)\hat N^m(\omega)
-\hat N^1(k-\omega) \hat V^m(\omega)\right)\,\omega\; d\omega \quad .
\end{eqnarray}
Modes that can contribute to the right hand side are the random
high frequency modes.

It is our belief that with sufficient randomness on high frequency
part thrown in, the right hand side of this equation will be
negligible upon the averaging over many nearly coincident strings.
For individual strings, each term in this equation must be roughly
of the same size. For a random collection of highly oscillatory
strings with arbitrary phase, it is likely that the averaging
procedure will introduce cancellations among nearby strings. It is
important to note that we are considering the larger scale motion
of these strings to be roughly the same. Only those variables
appearing on the right hand side are random.

For the sake of argument, we will proceed with this assumption.
Then, one is left with low frequency modes only. Performing
inverse Fourier transformation, we define
\begin{equation}
\tilde{n}^m= \frac{1}{2\pi}\int_{small}dk \;\hat N^m(k;\tau)
e^{ik\sigma},\qquad \tilde{v}^m= \frac{1}{2\pi}\int_{small}dk
\;\hat V^m(k;\tau) e^{ik\sigma}      .
\end{equation}
This sort of operation is what we mean by "averaging" or
"coarse-graining" the fundamental strings.

The averaged equation of motion for these slow variables
are simply (\ref{NGfluid})
except everything is replaced by tilde variables,
\bear
\p_0 \tilde n^m+\tilde v^1\p_1\tilde n^m
&=&\tilde n^1\p_1 \tilde v^m      ,\nonumber\\
\p_0\tilde v^m+\tilde v^1\p_1\tilde v^m &=&\tilde n^1\p_1\tilde
n^m      . \label{NGfluidtilde} \eear Now, the crucial difference
from the original equation of motion is that our tilde variables
need not saturate the constraints (\ref{NGconstraint}). This is
understandable because transverse fluctuations do not saturate all
degrees of freedom of our low energy theory, and the
energy-momentum have additional independent degree of freedom.
Interestingly, we can take an analogy at this point to the string
fluid coupled to tachyon matter, and introduce fictitious
'tachyonic' variables $\tilde n^T$ and $\tilde v^T$ to saturate
the constraints (\ref{NGconstraint}) or the energy-momentum of the
system,
\bear (\tilde n^T)^2+(\tilde v^T)^2&=&1-(\tilde n^1)^2
-(\tilde n^I)^2-(\tilde v^1)^2-(\tilde v^I)^2      ,\nonumber\\
\tilde n^T\tilde v^T&=&-\tilde n^1\tilde v^1 -\tilde n^I\tilde v^I
,\label{saturated} \eear
which may be solved
\bear &&
\tilde n^T=\frac{1}{2} \Bigg\{\sqrt{1-(\tilde n^I+\tilde v^I)^2-(\tilde
n^1+\tilde v^1)^2} +\sqrt{1-(\tilde n^I-\tilde v^I)^2
-(\tilde n^1-\tilde v^1)^2}\Bigg\}    \:  ,\nonumber\\
&&\tilde v^T=\frac{1}{2}\Bigg\{\sqrt{1-(\tilde n^I+\tilde v^I)^2
-(\tilde n^1+\tilde v^1)^2} -\sqrt{1-(\tilde n^I-\tilde
v^I)^2-(\tilde n^1-\tilde v^1)^2}\Bigg\}     \: .
\eear
An
intriguing fact is that they satisfy the exactly same form of
time-evolution equations as other ${\tilde n}^I$'s and
$\tilde{v}^I$'s do, \bear \p_0 \tilde n^T+\tilde v^1\p_1\tilde n^T
&=&\tilde n^1\p_1 \tilde v^T      ,\nonumber\\
\p_0\tilde v^T+\tilde v^1\p_1\tilde v^T &=&\tilde n^1\p_1\tilde
n^T      . \eear
which can be checked straightforwardly using
(\ref{NGfluidtilde}).

The above set of low-energy effective equations of motion for
Nambu-Goto string looks precisely same as those for string fluid
with tachyon matter, (\ref{fluidequation}). Although this already
provides a compelling hint at the identity of tachyon matter as an
effective degree of freedom representing small oscillations of
fundamental strings, we may go further and show the correspondence
more explicitly. Guided by the definitions of the fluid variables
in the string fluid side, let us define $\tilde \CH$ and $\tilde
\CP_1$ by \be \tilde n^1\,\,=\,\,\bigg\langle \frac{\cal
N}{H_{\cal N}}\bigg\rangle\,\,\equiv\,\,\frac{\CN}{\tilde \CH}
,\quad \tilde v^1\,\,=\,\,\bigg\langle \frac{P_{\cal N}}{H_{\cal
N}}\bigg\rangle\,\,\equiv\,\,\frac{\tilde \CP_1}{\tilde\CH}      .
\ee It follows that \be \tilde n^I\,\,=\,\,\tilde Y^I \bigg\langle
\frac{\cal N}{H_{\cal N}}\bigg\rangle\,\,=\,\, \frac{\CN\tilde
Y^I}{\tilde \CH} ,\quad \tilde v^I\,\,=\,\,\tilde\Pi_I
\bigg\langle \frac{\cal N}{H_{\cal N}}\bigg\rangle\,\,=\,\,
\frac{\CN\tilde \Pi_I}{\tilde \CH}      , \ee where $\tilde Y^I$
and $\tilde \Pi_I$ are slowly varying fluctuations. Furthermore,
we also introduce an effective field $\tilde T$ and its conjugate
$\tilde \pi_T$, which encode energy-momentum of small
oscillations, \be \tilde n^T\,\,\equiv\,\,\frac{\CN\p_1\tilde
T}{\tilde \CH} ,\quad \tilde v^T\,\,\equiv\,\,\frac{\tilde
\pi_T}{\tilde \CH} . \ee
Now the constraints (\ref{saturated})
read as
\be \tilde \CH \,\,=\,\, \sqrt{(\CN)^2+(\CN\tilde
Y^I)^2+(\tilde\pi_T)^2+(\tilde\CP_1)^2
+(\CN\tilde\Pi_I)^2+(\CN\p_1\tilde T)^2}    \quad  , \ee
and
\be \tilde
\CP_1+(\tilde\Pi_I)(\CN\tilde Y^I)+(\p_1\tilde T)(\tilde \pi_T)=0
. \ee

Noting the striking similarity of the above to the Hamiltonian
(\ref{fluidhamiltonian}) and momenta (\ref{fluidmomenta}) of the
string fluid system, we are ready to describe the mapping between
the two. Firstly, it is geometrically clear that $\CN\tilde
Y^I\,=\,\pi_1\tilde Y^I$ should be mapped to ${\cal P}_I$, observing that
$\tilde Y^I=\p_1 Z^I$. It is also obvious to identify the
conjugate momenta $\pi_I$ with $\CN\tilde \Pi_I$. Under these and
$\tilde T \rightarrow T$, the Hamiltonian and momenta match
precisely. We stress that this matching is not a mere artifact of
our definitions of $\tilde \CH$ and $\tilde \CP_1$, because the
fluid equations, which are identical on both systems, are non
trivial consequences of this Hamiltonian and the relevant
canonical structure.

This implies that we may consider $\tilde T \rightarrow T$ and
$\tilde \pi_T\rightarrow \pi_T$ be dynamical variables, instead of
being functions of $\tilde Y^I$ and $\tilde\Pi_I$. Once we assign the
canonical Poisson bracket,
\begin{equation}
\{T,\pi_T\}_{P.B.}=\delta ,
\end{equation}
also consider the Hamiltonian $\tilde \CH$ as a function of
$(D-2)+1$ pairs of conjugate variables, $\{Z^I,\pi_I;T,\pi_T\}$,
the evolution equation for the ${\tilde n}^T$, ${\tilde v}_T$ as
well as those for $\tilde H$, $\tilde\CP_1$, $\tilde n^I$, and
$\tilde v^I$ are all correctly reproduced. Poisson bracket for
$\pi_I$ and $Z^I$ may look as if they conflicted with those of
$\tilde \Pi_I$ and $\tilde Y^I$, due to the factor of $\cal N$.
Upon closer inspection, we must realize that the identification is
\begin{equation}
\pi_I=\sum_a \Pi_I^{(a)},
\qquad Z^I=\frac{1}{\cal N}\sum_a \tilde Y^I_{(a)} ,
\end{equation}
where the sum is over the ${\cal N}$ Nambu-Goto strings.
This produces the expected Poisson bracket for the
averaged variables. Thus, we managed to
construct an effective scalar field $T$ on a coarse-grained
Nambu-Goto string, which captures the energy-momentum
content of microscopic part of the string oscillation.

\section{Beyond Classical Limit}

We have provided several arguments supporting the claim that
string fluid with tachyon matter describes a macroscopic bundle of
fundamental strings. What this means in the larger issue of fully
quantum closed strings in the open string tachyon condensation
remains to be seen. In a way, the above example of macroscopic
interpretation in fully classical limit is itself quite surprising
and may lead to hitherto unknown relationship between closed
strings and open strings.

At the same time, we are again tempted to ask questions about
whether and how some part of this story with classical story is
lifted to the fully quantum level. Recent progresses found a
fertile ground for asking such a fully quantum mechanical
question, namely old matrix models reinterpreted as a theory of an
open string tachyon
\cite{McGreevy:2003kb,Klebanov:2003km,McGreevy:2003ep,Takayanagi:2003sm,Douglas:2003up,Martinec:2003ka,Sen:2003iv,Gutperle:2003ij}.
In this simplified setting of two spacetime dimensions, closed
strings emerge as a collective excitation of the open string
tachyon theory in the fully non perturbative definition of the
latter.

One possible scenario of how fundamental strings would show up in
the quantum open string theory in general dimensions was given
some time ago \cite{Bergman:2000xf,Yi:1999hd}. It remains to be
seen whether one could find a common ground among the fully
quantum mechanical case (but limited to 2-spacetime dimensions), a
low energy viewpoint of the present note, and the somewhat
pictorial but fully quantum mechanical scenario.

\vskip 1cm \centerline{\large \bf Acknowledgement} \vskip 0.5cm
P.Y. thanks K. Hori, M. Rozali, A. Sen, and L. Susskind for
helpful conversations. H.-U.Y. is grateful to Yoonbai Kim for
discussions. H.-U.Y. is partly supported by grant
No.R01-2003-000-10391-0 from the Basic Research Program of the
Korea Science \& Engineering Foundation. P.Y. is supported in part
by Korea Research Foundation (KRF) Grant KRF-2002-070-C00022.

\end{document}